\documentclass[aps, prd, onecolumn, amsmath, superscriptaddress, floatfix, nofootinbib, preprintnumbers, preprint]{revtex4-1}

\usepackage{graphicx}
\usepackage{dcolumn}
\usepackage[colorlinks=true, urlcolor=blue, linkcolor=blue, citecolor=blue]{hyperref}
\usepackage{comment}
\usepackage{xcolor}
\usepackage{orcidlink}
\usepackage[percent]{overpic}
\usepackage{soul}

\newcommand{\Ocal}{\mathcal{O}}

\begin{document}

\title{White Paper on Polarized Target Studies with Real Photons in Hall D}

\author{F.~Afzal\,\orcidlink{0000-0001-8063-6719}}
\affiliation{Helmholtz Institute for Radiation and Nuclear Physics, University of Bonn, Germany}
\author{M. M.~Dalton\footnote{Contact: \href{mailto:dalton@jlab.org}{dalton@jlab.org}}\,\orcidlink{0000-0001-9204-7559}}
\affiliation{Thomas Jefferson National Accelerator Facility, Newport News, USA}
\author{A.~Deur\,\orcidlink{0000-0002-2203-7723}}
\affiliation{Thomas Jefferson National Accelerator Facility, Newport News, USA}
\author{P.~Hurck\,\orcidlink{0000-0002-8473-1470}}
\affiliation{University of Glasgow, United Kingdom}
\author{C.D.~Keith\,\orcidlink{0000-0002-5467-2983}}
\affiliation{Thomas Jefferson National Accelerator Facility, Newport News, USA}
\author{V.~Mathieu\,\orcidlink{0000-0003-4955-3311}}
\affiliation{Departament de F\'isica Qu\`antica i Astrof\'isica and Institut de Ci\`encies del Cosmos,\\ Universitat de Barcelona, E-08028, Spain}
\author{S.~\v{S}irca\,\orcidlink{0000-0002-9962-9663}}
\affiliation{Faculty of Mathematics and Physics, University of Ljubljana, Ljubljana, Slovenia}
\author{Z.~Yu\,\orcidlink{0000-0003-1503-5364}}
\affiliation{Thomas Jefferson National Accelerator Facility, Newport News, USA}

\date{\today}
\preprint{JLAB-PHY-24-4101}

\begin{abstract}
This white paper summarizes the Workshop on Polarized Target Studies with Real Photons in Hall D at Jefferson Lab, that took place on 21 February 2024.
The Workshop included about 45 participants both online and in person at Florida State University in Tallahassee.  
Contributions describe the experimental infrastructure available in Hall D and potential physics applications.
The rate and detection capabilities of Hall D are outlined, as well as the properties of a circularly polarized photon beam and a polarized target.
Possible physics measurements include light and strange quark baryon spectroscopy, the GDH sum rule, proton structure accessed through measurement of Generalized Parton Distributions and modification of nucleon structure within the nuclear medium.

\end{abstract}

\maketitle

\tableofcontents

\section{Introduction}

Hall D at Jefferson Lab is a multipurpose facility with a  high energy photon beam and largely hermetic detector~\cite{GlueX:2020idb}.  
Experiments in Hall D have used unpolarized targets including hydrogen, deuterium, helium, beryllium, carbon and lead, for wide variety of different physics goals.
The photon beam has typically had a linearly polarized component at energies around 9~GeV, from coherent bremsstrahlung, and recently data with circular polarization has also been accumulated.
What has been absent thus far is any polarization of the target.
In this paper we explore some of the physics opportunities that would be available with a polarized target in Hall D.

\section{Capabilities of Hall D (Mark Dalton)}

Hall D infrastructure consists of the Hall D experimental hall with large solenoidal spectrometer with high efficiency for both charged and neutral particles.  
The typical momentum resolution for charged particles is 1--3\%, while the resolution is 8--9\% for very-forward high-momentum particles.
A detailed description of the apparatus is given in Ref.~\cite{GlueX:2020idb}.

The photon beam is produced in the Tagger Hall from  bremsstrahlung (typically coherent from an aligned diamond radiator).
The energy and arrival time of beam photons are tagged by detecting the production electron using a large dipole tagging magnet, a scintillator hodoscope and a scintillating fiber array.
The photon beam energy is determined with a resolution of 0.1--0.3\% depending on the energy.
For exclusive final states a kinematic fit allows the excellent beam energy resolution to improve the final state resolution. 
The photon flux is determined using a pair spectrometer, while the linear polarization of the photon beam is determined using a polarimeter based on triplet photoproduction.
The circular polarization is determined by measuring the electron beam polarization either before the beam acceleration or in one of the three other experimental halls and studying the precession of the beam spin through the accelerator.
If necessary, additional dedicated time could be used to help constrain the electron beam longitudinal polarization to $<2\%$ if it has a favorable angle arriving in the hall.

For GlueX II running, the detector operates routinely at trigger rates of 70~kHz and data rates of 1.2 gigabytes per second. 
This is achieved with a 300~nA electron beam current on a 50~$\mu m$ diamond wafer producing photons incident on a 30~cm long liquid hydrogen (LH$_2$) target, 3.4\% of a radiation length.
In this condition, only photons with energy 8~GeV and higher can be tagged.
In order to run the tagger down to 3 GeV the beam current would need to be decreased to 100~nA as was done during GlueX I Low Intensity (LI) in 2016 and 2017.
For GlueX III it is planned to again double the luminosity, by operating at 600 nA.
In order to keep similar trigger and data rates, the trigger will be modified, likely by increasing the energy threshold.
In the  present configuration, the rate will then be limited by hodoscopes in the tagger system and the most inner paddles in the time of flight system, which begin to suffer unmanageable pileup.

For the purpose of estimating rates from a polarized target, the target is expected to easily maintain a long polarization lifetime at any flux that can be reasonably be handled by the Hall D setup.
A 10 cm butanol target has 3.1 times as many nucleons per unit area as the regular 30 cm LH$_2$ target.
Thus, a similar luminosity to GlueX III can be achieved by decreasing the photon beam flux by a factor of 1.55 from GlueX II (by operating with  a 190 nA electron beam current).
The polarized target will be about 17.8\% of a radiation length, so the electromagnetic rate in the trigger and electromagnetic load on the FCAL and FDC  will increase by a factor of 5 for the beam flux, which may limit the maximum luminosity.  Hall D has previously operated with a carbon target of 7.9\% at 150 $\mu A$.

The polarized protons are a fraction $10/74=0.135$ of the total nucleons.
The number of polarized protons per unit area is 0.42 times as many as protons in the regular LH$_2$ target.
Thus, at the GlueX III luminosity, the polarized proton luminosity will be 0.27 that of GlueX II and will still allow tagging down to about 5~GeV.  Tagging down to 3 GeV would require lowering the beam current.
This is summarized in Table~\ref{tab:lumi}.

Background stemming from non-target materials, e.g., beamline windows and air, should be a smaller fraction of the total trigger rate.
With the  LH$_2$, the trigger rate from non-target interactions is 1/4 of the trigger rate and for L$^4$He in the same target volume, it  is 1/5 of the trigger rate.  
For the polarized target, the non-target fraction will be smaller still.

It is also possible to move the coherent peak to different energies, or to  produce a different bremsstrahlung spectrum using a different radiator, which will change the simple scaling presented here.

\begin{table}[!ht]
\caption{{\small Table shows the luminosity for various phases of GlueX running and a possible configuration for the polarized butanol target which would give the same luminosiyty as the proposed GlueX III.
These numbers are approximate and assume a coherent bremsstrahlung beam with peak at 9~GeV.  
LI refers to a lower intensity running at the start of GlueX I.  
The photon energy range that can feasibly be tagged  depends on the beam current (assuming constant radiator thickness). 
}}
\centering
\begin{tabular}{l|c|c|c|c|c|c}
 & tagging &  beam    & \multicolumn{2}{c|}{target nucleons} &  \multicolumn{2}{c}{relative lumi}\\
 & range &  current  & total   & signal  & total   & signal  \\
 & GeV & nA &  $\times10^{24}\textrm{ cm}^{-2}$  & $\times10^{24} \textrm{ cm}^{-2}$  & \\
\hline
GlueX I LI       & $3-12$ & 100 & 1.27 & 1.27 & 0.3 & 0.3  \\
GlueX I          & $5-12$ & 150 & 1.27 & 1.27 & 0.5 & 0.5  \\
GlueX II         & $8-12$ & 300 & 1.27 & 1.27 & 1.0 & 1.0  \\
GlueX III        & $8-12$ & 600 & 1.27 & 1.27 & 2.0 & 2.0  \\
Polarized butanol & $5-12$ & 190 & 3.95 & 0.53 & 2.0 & 0.27 \\ 
\hline
\end{tabular}
\label{tab:lumi}
\end{table}

\section{Overview of studies with circular polarization at GlueX (Peter Hurck)}
\label{sec:circpol}
Many of the topics covered below require not only a polarized target but also a circularly polarized photon beam. Recently a proposal was submitted to PAC52 that asks for longitudinal electron polarization to be delivered to Hall D~\cite{Hurck:2024alu}. This will enable GlueX to run with circularly and linearly polarized photon beams. \par
The main motivation for the proposal is the measurement of the weak decay constant $\alpha_-$ for $\Lambda\rightarrow p\pi^-$. The measurement will exploit the fact that having both circular and linear polarization simultaneously in the beam results in an over-constrained set of reaction amplitudes that describe the production of $\Lambda$ baryons. This will allow to leave $\alpha_-$ as a free parameter in the fit. All details of this measurement can be found in Ref.~\cite{Hurck:2024alu}. \par
An important aspect of the proposed measurement is that it requires both circular and linear photon beam polarization, sometimes referred to elliptical polarization. 
Elliptical polarization is produced by having a longitudinally polarized electron beam produce coherent bremsstrahlung on a thin diamond radiator. 
This has been successfully demonstrated in previous measurements~\cite{A2:2024ydg}. 
The resulting circular component of the beam will require a small correction to the well known relationship shown in Eq.~\eqref{eq:circpol}, where $P^\gamma_C$ denotes the degree of circular photon beam polarization, $p_e$ denotes the degree of longitudinal electron beam polarization, and $x=E_\gamma/E_e$ is the ratio of photon to electron beam energy.
\begin{equation}
    \label{eq:circpol}
    P^\gamma_C = p_e(4x-x^2)/(4-4x+3x^2).
\end{equation}
\begin{figure}
    \centering
    \includegraphics[width=0.8\linewidth]{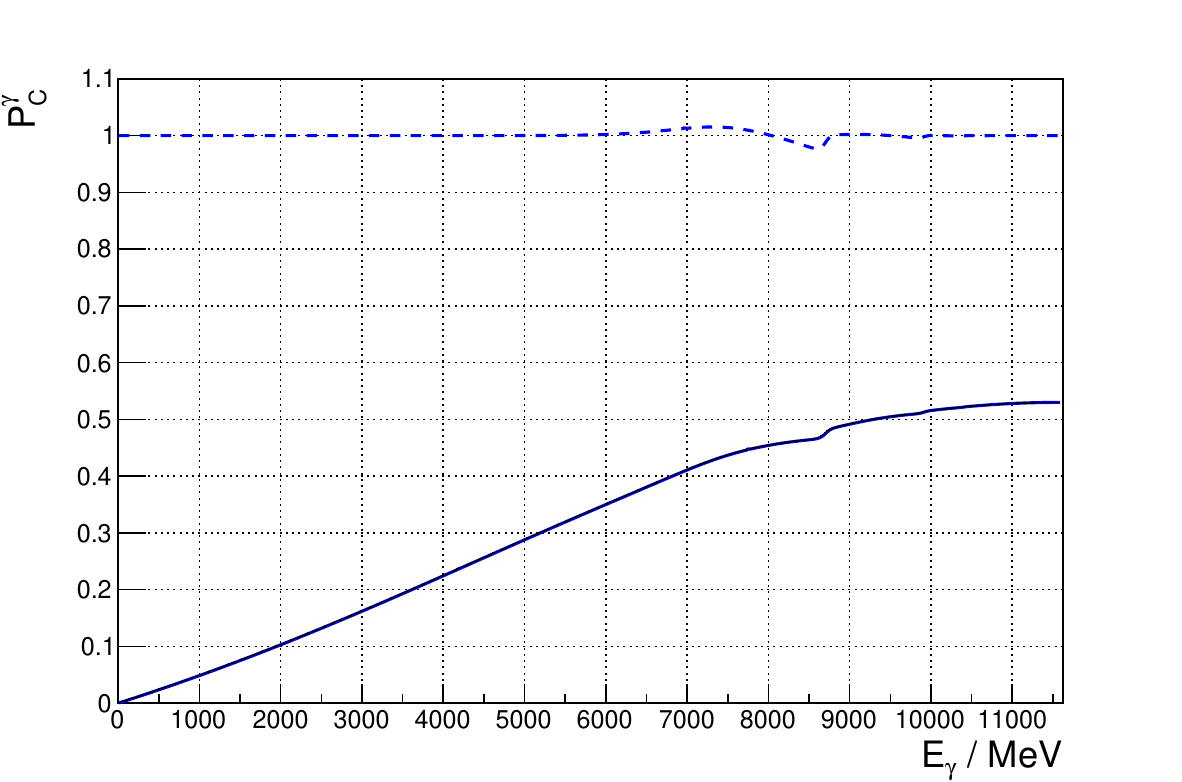}
    \caption{The degree of circular photon beam polarization as a function of the photon energy when using longitudinally polarized electrons on a diamond radiator (solid line). The relative deviation between the solid curve and Eq.~\eqref{eq:circpol} are shown with the dashed line at the top. Relevant beam, collimator and diamond parameters are matched to the GlueX data taking in 2023 with $p_{e}=0.53$. Taken from Ref.~\cite{Hurck:2024alu}.}
    \label{fig:pcirc}
\end{figure}
The effect of this correction is visualized in Figure~\ref{fig:pcirc}. At standard GlueX running conditions it is only a small. However, it will become more important if the coherent peak position is moved to lower photon beam energies. \par
Having an elliptically polarized photon beam also opens the door to a variety of other measurements. The circular component of the photon beam would give access to the third component of spin-density matrix elements, which cannot be measured in standard GlueX running. This could provide further information on the production process of a reaction. In the same manner, one could also use the circular polarization information to constrain amplitude models for the GlueX meson spectroscopy program. This could help to reduce ambiguities in the results and help the core GlueX mission. Circular polarized photons are also needed to study the helicity asymmetry in Timelike Compton Scattering which provides important inputs for Generalized Parton Distributions.

\section{Polarized Target Capabilities (Chris Keith)} \label{sec:poltar}
The target will be polarized by the technique of dynamic nuclear polarization
(DNP) which has previously been utilized in the other three experimental halls at Jefferson Lab~\cite{Keith:2016/V}.  The operating conditions in Hall D will be a temperature of approximately 0.3~K and a magnetic field of 2.5~T, where one can reasonably expect a proton and deuteron polarizations of 90\% and 70\%, respectively. A number of other nuclei have been successfully polarized using DNP.  The COMPASS collaboration has demonstrated $^6$Li and $^7$Li polarizations of 50\% and 90\% in irradiated samples of LiD~\cite{BALL2003101}.  Using paramagnetic radicals with extremely narrow EPR lines, Meyer {\em et al.} demonstrated $^{13}$C polarizations up to 75\% at 0.9~K and 5~T~\cite{MEYER20111}.  We expect similar or even higher polarizations under our anticipated temperature and field conditions.  A $^{19}$F 
polarization of 80\% was achieved at 0.4~K and 2.5~T
in a fluorinated alcohol doped with the Cr(V) paramagnetic complex~\cite{Hill1989}.  High polarizations may also be possible in $^{17}$O.
Michaelis {\em et al.} reported a 80-fold enhancement in the polarization of $^{17}$O in enriched H$_2$$^{17}$OH samples at 80~K and 5~T~\cite{Michaelis}.  A similar enhancement at 0.3~K and 2.5~T corresponds to a polarization of approximately 50\%.  

The target samples will be cooled using a horizontal $^3$He-$^4$He dilution refrigerator similar to the one described by Keith {\em et al}~\cite{Keith201227}, and the magnetic field will be generated by the Hall D superconducting solenoid.  For optimal dynamic nuclear polarization, one normally aims for a field homogeneity $\Delta B/B$ of about 100~ppm over the volume of the targets sample. Unfortunately, the Hall D solenoid does not meet this requirement for our assumed 10~cm sample length, nor does it reach 2.5~T.  We therefore intend to incorporate a series of thin, superconducting coils inside the target cryostat that will simultaneously increase the field strength and improve its homogeneity to the desired values.  Initial studies find that three coils, operated with independent currents, can successfully increase the on-axis field to 2.5~T and improve its uniformity to less than 50~ppm \cite{Lagerquist}.  The winding parameters of the coils is given in Table~\ref{tab:Shims}.

For most applications it will be sufficient to continuously polarize the target samples using 70~GHz microwaves and a sample temperature of 0.2--0.3~K.  However, there are situations in which a frozen-spin operation of the target is beneficial.  Here, the polarized sample is cooled to an ultralow temperature ($<50$~mK) when the microwaves are switched off, and scattering data are acquired while the polarization slowly decays in an exponential manner,
\begin{equation}
    P(t) = P_{te} + P(0)e^{-t/T_1}, 
\end{equation}
where $P_{te}$ is the thermal equilibrium polarization of the nuclei at the operational field and temperature conditions, and $T_1$ is the spin-lattice relaxation time of the nuclei at those conditions. The Hall B frozen spin target demonstrated a proton $T_1 = 3000-4000$~h at 30~mK and 0.56~T, and the frozen spin target at Mainz achieved $T_1 = 1500$~h for deuterons at 25~mK and 0.45~T~\cite{Thomas}.  The relaxation at 2.5~T should be considerably slower.  

Operating the target in frozen-spin mode can be very advantageous for studying tensor polarized deuteron observables.  The tensor polarization of spin-1 nuclei can be written
\begin{equation}
\label{eqn:Pzz}
P_{zz} = 1 - 3N_0,
\end{equation}
where $N_0$ is the fraction of deuterons in the $m=0$ magnetic substate.
While DNP can produce vector polarizations exceeding $\pm$70\% and positive tensor polarizations of $+40$\%, it cannot generate {\em negative} tensor polarizations, meaning it cannot enhance the $m=0$ population above the unpolarized value of $1/3$. However, our initial estimates indicate that negative values of $P_{zz}$ approaching -70\% can be obtained in the frozen spin state using the standard NMR technique of Adiabatic Fast Passage~\cite{Dalton}.

\begin{table*}[]
\label{tab:Shims}
    \centering
    \begin{tabular}{|c|c|c|c|c|c|c|}
    \hline
        Target & Coil &  Coil & Coil & Windings & Number of & Coil\\
        Center (cm) & Number & Center (cm) & Length (cm)  & per Layer & Layers & Current (A) \\
        \hline
        \hline
         130 & 1 & -3.3 &  22.9 & 996  & 4 & 36.3 \\
            & 2 & -1.2 &  14.2 & 617 & 2 & -11.5 \\
            & 3 & 1.1  &  6.4 & 278.3 & 1 & -1.0\\
         \hline    
    \end{tabular}
    \caption{Preliminary winding parameters for three correction coils assuming a central target location of 130~cm relative to the upstream edge of the Hall D solenoid.  Coil center is measured relative to the target's center position.  A negative current produces a magnetic field antiparallel to the solenoid's field.  The wire diameter is 0.23~mm, and the coil diameter is 2.0 cm.}
\end{table*}

\section{REGGE: High-energy contribution to the GDH sum rule (Simon \v{S}irca)}
\label{sec:REGGE}

The Real Gamma GDH Experiment (REGGE)~\cite{Dalton:2020wdv} is an approved Hall~D experiment devoted
to the measurement of the high-energy contribution to the 
Gerasimov--Drell-Hearn (GDH) sum rule for the proton and the neutron.   
This is a general relation that is valid for any type of particle
and links the anomalous magnetic moment $\kappa$ of a particle 
to its helicity-dependent photoproduction cross-sections:
\begin{equation}
I \equiv \int_{\nu_0}^{\infty}\frac{\Delta \sigma(\nu)}{\nu}\,\mathrm{d}\nu=\frac{4\pi^2 S \alpha_\text{em}\kappa^2}{M^2} \>.
\label{eq:GDH}
\end{equation}
Here $\nu$ is the probing photon energy, $S$ is the spin of 
the target particle, $M$ is its mass, and $\nu_0 = m_\pi(1+m_\pi/2M)$ 
is the threshold energy for pion photoproduction.
In the case of a proton or neutron target  $S=1/2$ and 
$\Delta \sigma \equiv \sigma_{P} - \sigma_{A}$
is the difference in total photoproduction cross-sections 
($\gamma N \to X$) for which the  photon spin is parallel 
and anti-parallel to the target particle spin, respectively. 

Although the GDH sum rule is based on very fundamental quantum field
theory assumptions, there are reasons to be concerned.  The 
{\sl unpolarized\/} total photoabsorption cross-section
$\sigma_P + \sigma_A$ increases with energy as $\sim\nu^{0.08}$
at high energies --- at least up to $\nu\approx 10^4\,\mathrm{GeV}$!
It is not clear what consequence this divergence in the unpolarized 
cross-section implies for the polarized cross-section difference 
and for the convergence of the GDH integral.
Moreover, several theoretical studies, some more speculative than
others, have been presented that indicate why and how the GDH sum rule 
itself might be modified.  

\begin{figure}[!hbtp]
\centering
\includegraphics[width=0.75\textwidth]{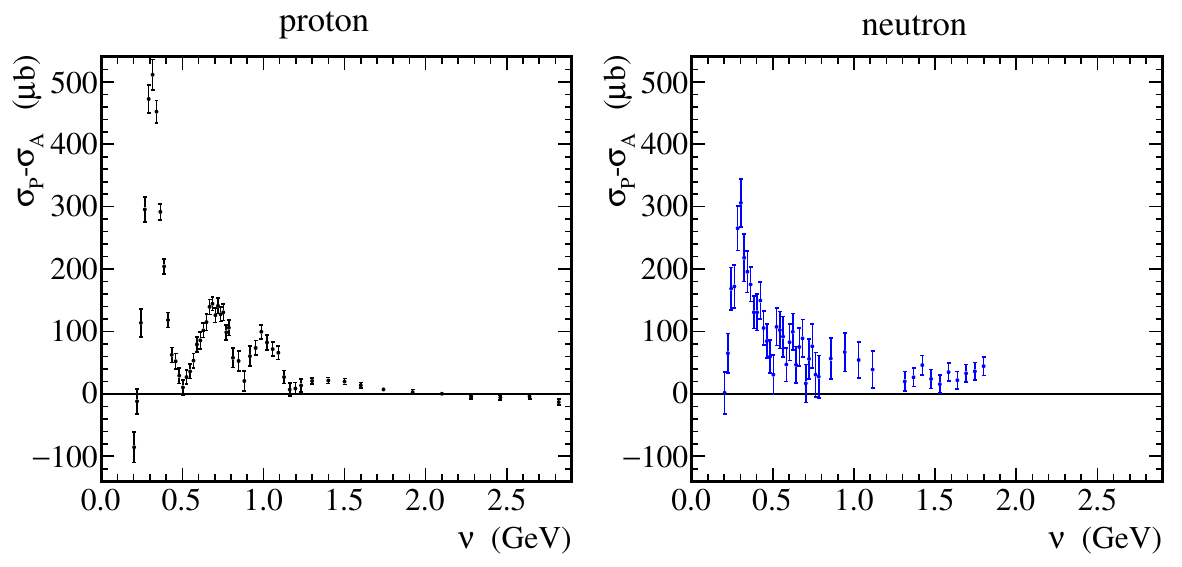}
\vspace{-0.3cm}
\caption{\label{fig:GDH_world}World data of the spin-dependent cross-section difference $\Delta \sigma$ on the proton (left) and neutron (right).  Data from various experiments have been combined and rebinned.}
\end{figure}

The primary motivation of the REGGE experiment is to explore 
the integrand $\Delta\sigma(\nu)$ in the high-energy domain, 
up to the values of $\nu$ accessible at JLab.  It is very
desirable to perform this task as no data for $\Delta\sigma(\nu)$ 
exist beyond $\sim 3\,\mathrm{GeV}$ for the proton and
$\sim 2\,\mathrm{GeV}$ for the neutron; see Fig.~\ref{fig:GDH_world}.

\begin{figure}[!hbtp]
\centering
\includegraphics[width=0.75\textwidth]{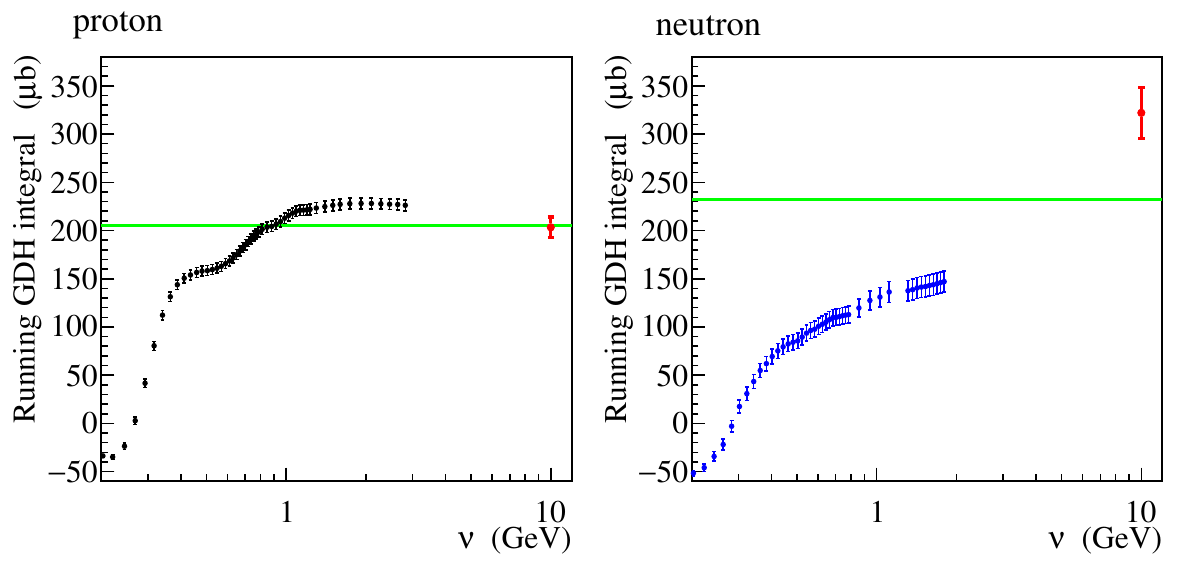}
\vspace{-0.3cm}
\caption{\label{fig:GDHrun_pn}The ``running'' GDH integral 
for the proton (left) and neutron (right) starting at $\nu=0.2$\,GeV,
the lowest experimentally attained energy.
The $\nu_0 \le \nu\le 0.2\,\mathrm{GeV}$ contributions are estimated to be
$(-28.5 \pm 2)\,\mu\mathrm{b}$ and $\approx -41\,\mu\mathrm{b}$,
respectively.  The green horizontal lines show the expected value 
according to Eq.~(\ref{eq:GDH}).  The red points are the recent 
generalized GDH results from electroproduction extrapolated to $Q^2=0$.}
\end{figure} 

Furthermore, the ``running'' GDH integral, that is, the integral
(\ref{eq:GDH}) with the running upper integration bound, does not
seem to saturate if only the existing data are taken into consideration,
regardless of the fact that $\Delta\sigma(\nu)$ is  weighted by
$1/\nu$ in the integral; see Fig.~\ref{fig:GDHrun_pn}.  
Extrapolation is needed to account for the uncovered energy range, 
and usually one resorts to Regge phenomenology to do this.
In Regge formalism, the polarized cross-section difference 
has a very simple form
$$
\Delta\sigma = Ic_1 s^{\alpha_{a_1}-1} +c_2 s^{\alpha_{f_1}-1} \>,
$$
where $s=2M\nu + M^2$, $I=\pm$ is the isospin sign of the proton 
or neutron, $c_1$ and $c_2$ are well-known constants, while 
$\alpha_{a_1}$ and $\alpha_{f_1}$ are the parameters describing 
the intercepts of the $a_1$ and $f_1$ Regge trajectories.
Measuring $\Delta\sigma(\nu)$ to $\sim 12\,\mathrm{GeV}$ will
allow us to determine these intercepts to high precision and
help us better to understand both the behavior of the integrand
and the convergence of the integral.  While this may not be
considered as a ``flagship'' effort, it must be considered
as an imperative benchmark for all Hall~D doubly-polarized efforts:
colloquially speaking, if one cannot accomplish REGGE, 
one cannot accomplish anything.

\begin{figure}
\center 
\includegraphics[height=8cm]{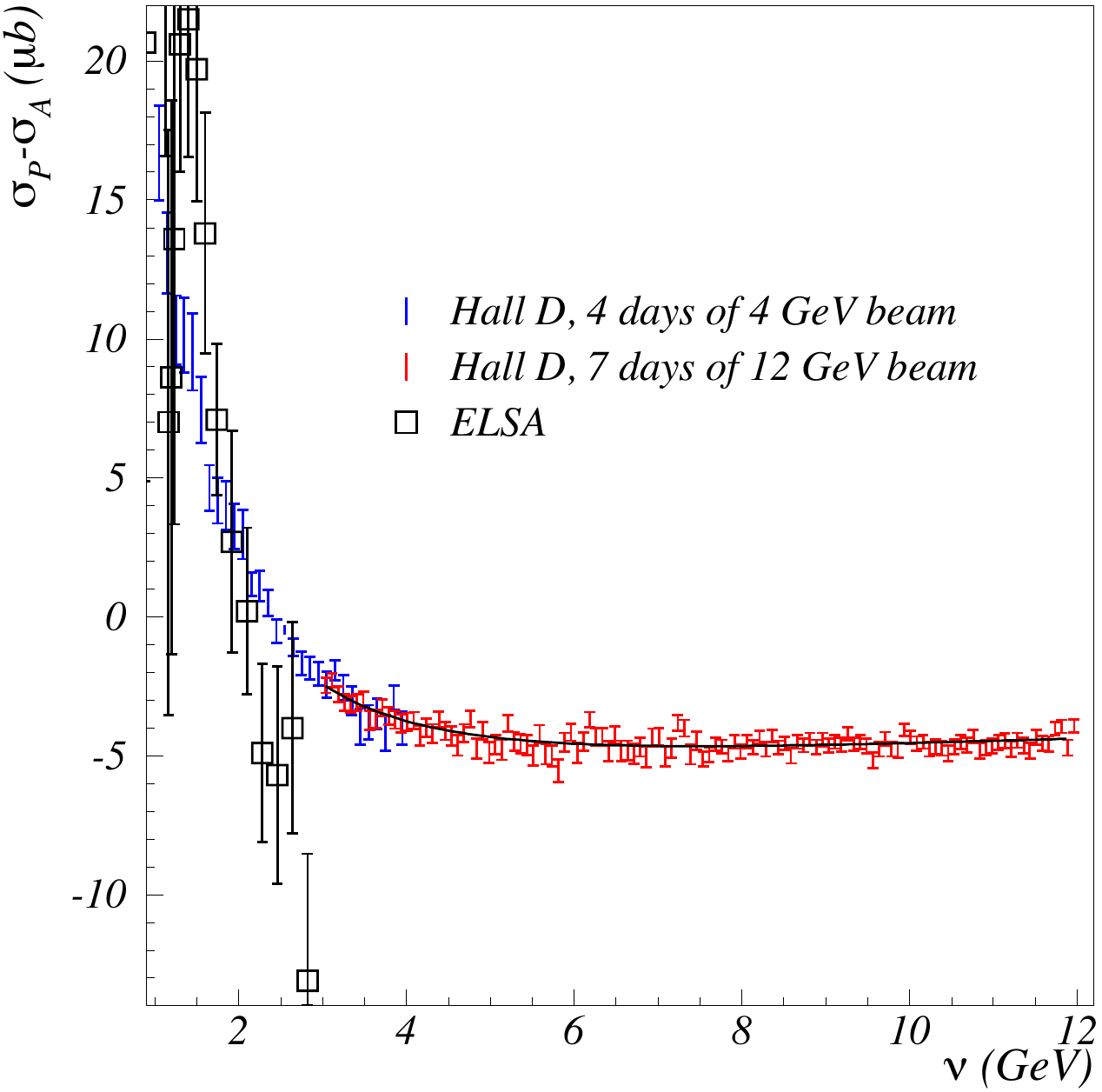}~~~%
\includegraphics[height=8cm]{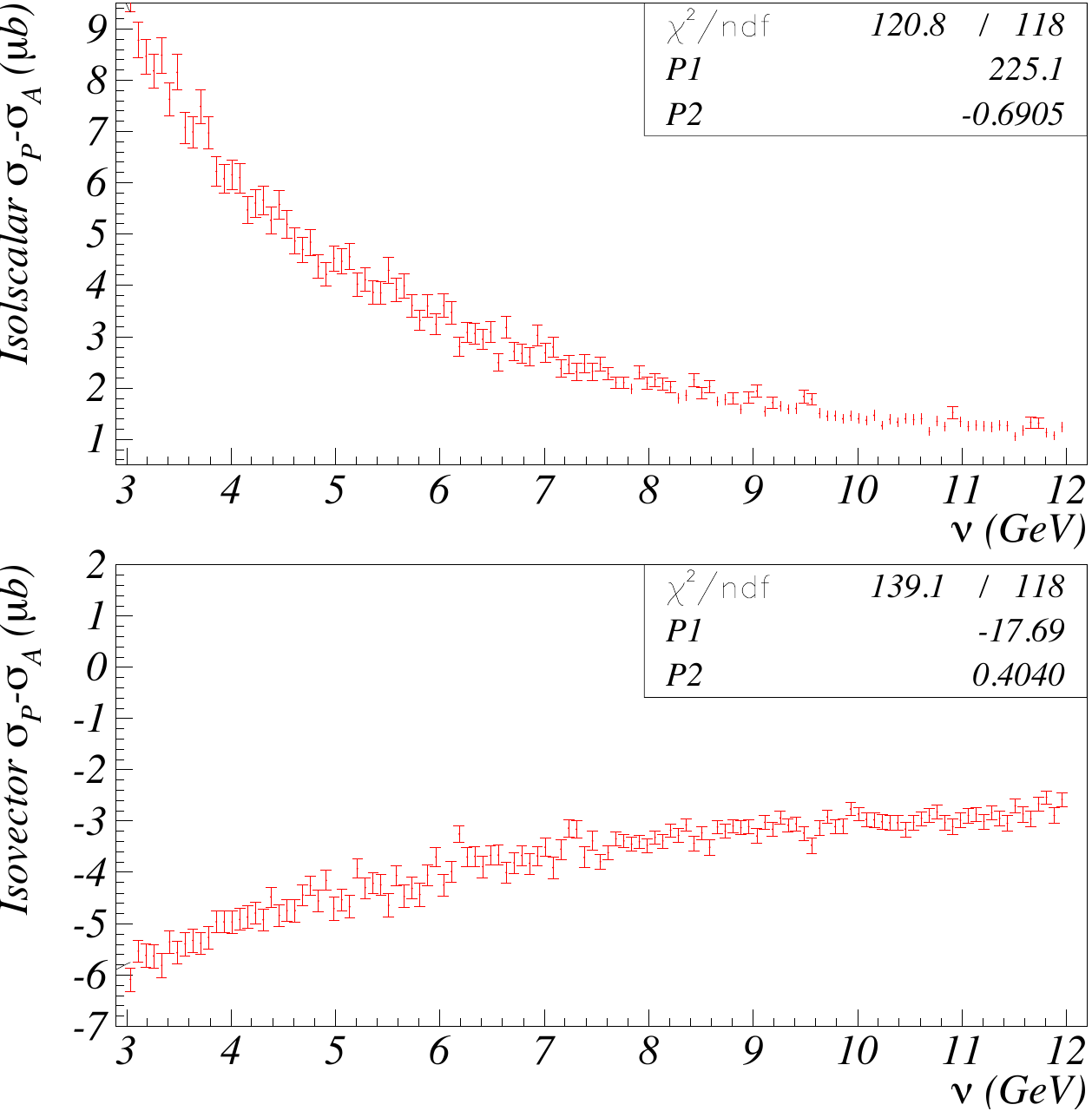}
\caption{\label{fig:GDHpISO}Left: $\Delta\sigma$ on the proton 
from ELSA high-$\nu$ data (squares) and expected results from Hall D 
using a 12 GeV beam (red) and a 4 GeV beam (blue). 
Right: Isospin decomposition of  $\Delta \sigma$.}
\end{figure}

The experiment will require a circularly polarized photon beam 
(produced from a longitudinally polarized electron beam) with 
a photon flux of about $0.7\cdot 10^7$/s.  The experiment will run 
in two configurations which require two different CEBAF beam energies,
12 GeV and 4 GeV.  A new longitudinal polarized proton and deuteron target 
 will be needed.  The experiment has been approved for 21 PAC days 
at 12 GeV and 12 PAC days at 4 GeV.  The expected precision of the
results on $\Delta\sigma$ for the proton and for the isospin decomposition 
of $\Delta\sigma$ is shown in Fig.~\ref{fig:GDHpISO}.

\section{Medium modification of the nucleon spin structure (Alexandre Deur)}
\label{sec:REGGEON}

Recent studies on nucleon short-range correlations (SRC) in nuclei~\cite{Hen:2013oha, Hen:2016kwk} have stimulated a lively debate on the origin of the EMC effect~\cite{EuropeanMuon:1983wih}. The two leading competing scenarios for such origin are the mean field (MF) mechanism~\cite{Thomas:1989vt, CiofidegliAtti:2015lcu} and the SRC scenario~\cite{Frankfurt:1988nt}. It is expected that polarized data with virtual~\cite{Thomas:2018kcx} and real photons~\cite{Bass:2020bkl, Bass:2022pyx} can crucially help settling the debate. In particular, it is clear that for a nucleon embedded in nuclear matter, both the integral and the integrand of the GDH sum rule (Eq.~(\ref{eq:GDH}) in Section~\ref{sec:REGGE}) will be significantly modified by the nuclear medium. In a Letter of Intent for PAC 51, the Real Gamma GDH Experiment on Nuclei (REGGEON)~\cite{REGGEON-loi} proposes to measure in Hall D the GDH integrand, $\Delta \sigma/\nu$, for light to medium-sized polarized nuclei with circularly polarized photons in the energy range of approximately 1.5--12\,GeV.

The expectation that GDH for an embedded nucleon is sensitive to medium modification arises from distinct but consistent considerations~\cite{Bass:2020bkl, Bass:2022pyx}. Firstly, the static part of the GDH sum rule (RHS of Eq.~(\ref{eq:GDH}) is proportional to $\kappa/M$, the ratio of the anomalous magnetic moment to the mass of a nucleon, with both $\kappa$ and $M$ expected to be {\it differently} affected by medium modifications: $\kappa$ is predicted to increase and $M$ to decrease, making the imbedded nucleon GDH sum rule value higher than that of a free nucleon. Secondly, the GDH integrand (LHS of Eq.~(\ref{eq:GDH}) is dominated by contributions  from resonances whose masses shift in nuclear matter. Therefore, due to the $1/\nu$ weighting of $\Delta \sigma$, a resonance contribution is suppressed or strengthened if its in-medium mass increases or decreases, respectively. Evidences indicate that the $\Delta$ is lighter in nuclear matter and since it dominates the GDH integral, one expects an increase of the GDH sum rule, consistent with the $\kappa/M$ expected modification. Other effects will contribute to the medium modification of GDH, such as resonance width broadening and EMC-like structure modifications in the high-$\nu$ domain. In all, the modification is should be sizeable: for example, the GDH sum rule for a proton embedded in $^7$Li should be on the order of $\sim$270~$\mu$b, an increase of $\sim$25\% compared with 204.78~$\mu$b for a free proton.

Measuring the modifications of $\kappa/M$ in a nucleus would be a new and unique contribution to the general study of the modification of nucleon properties in nuclear matter. However, the envisioned 1.5--12\,GeV energy range of the integral will not cover the low-$\nu$ contribution to GDH. Therefore, a complementary measurement at a lower energy facility is desirable.

Independently of the ability to measure the full GDH integral, its high energy part is should be also significantly sensitive to medium modification: comparing the modification anticipated for the static side of the GDH sum rule, viz on $\kappa/M$, and the resonance modification on the integral side indicates that more than 40\% of the medium modification would come from the high-$\nu$ domain~\cite{REGGEON-loi}. This, together with the suppressing $1/\nu$-weighting, suggests a large medium modification of $\Delta \sigma$ in the high-$\nu$ domain. Its measurement can then be directly and precisely compared to that from REGGE in Section~\ref{sec:REGGE} in order to provide for the first time the medium modification of the Regge intercept and slope controlling the high-$\nu$ dependence of the polarized photoproduction cross-section. Ref.~\cite{REGGEON-loi} estimated that 11 weeks of running on four different nuclei ($^7$Li, $^{13}$C, $^{17}$O, and $^{19}$F) would be sufficient for such measurement, with the different nuclei allowing to establish how the modification depends isospin and on nuclear matter density.

Another goal of REGGEON is to measure beam-target double-polarization asymmetries for several types of mesons photoproduced with a correlated nucleon pair. The dependence of such asymmetries on the nucleon relative momentum probes the polarization of the nucleons in the correlated pair, which is predicted to be suppressed in neutron-proton SRC pairs~\cite{Thomas:2018kcx} but not in the MF scenario. Therefore, such a signal has the potential to settle the question of the origin of the EMC effect. The asymmetry in the case of $^7$Li should range from a few \% to 10\%, thus being measurable within a few weeks of running. Ref.~\cite{REGGEON-loi} estimated 16 weeks for a 2\% precision on the asymmetry for four energy bins.

REGGEON does not require any new instrumentation beyond that already discussed in this White Paper. The proposed target nuclei span from light to medium mass nuclei and are $^7$Li, $^{13}$C, $^{17}$O, and $^{19}$F. The DNP target envisioned for REGGE (Section~\ref{sec:poltar}) can straightforwardly accommodate these nuclei. To cover the proposed energy range of 1.5--12\,GeV necessitates two different electron beam energies.

\section{Exclusive photoproduction reactions for nucleon structure through generalized parton distributions (Zhite Yu)}

%
%

\newcommand{\fig}[1]{Fig.~\ref{#1}}
\newcommand{\eq}[1]{Eq.~\eqref{#1}}

\newcommand{\pp}[1]{\left(#1\right)}
\newcommand{\bb}[1]{\left[#1\right]}
\newcommand{\cc}[1]{\left\{#1\right\}}
\newcommand{\vv}[1]{\langle #1 \rangle}

\newcommand{\beq}[1][]{\begin{equation}\label{#1}}
\newcommand{\eeq}{\end{equation}}

\newcommand{\nn}{\nonumber}

\newcommand{\wt}[1]{\widetilde{#1}}
\newcommand{\M}{\mathcal{M}}
\newcommand{\Mt}{\wt{\mathcal{M}}}
\newcommand{\N}{\mathcal{N}}
\newcommand{\F}{\mathcal{F}}
\renewcommand{\H}{\mathcal{H}}
\newcommand{\Ft}{\widetilde{\mathcal{F}}}

\begin{figure}[htbp]
	\centering
		\includegraphics[trim={0 -6em 0 0}, clip,scale=0.45]{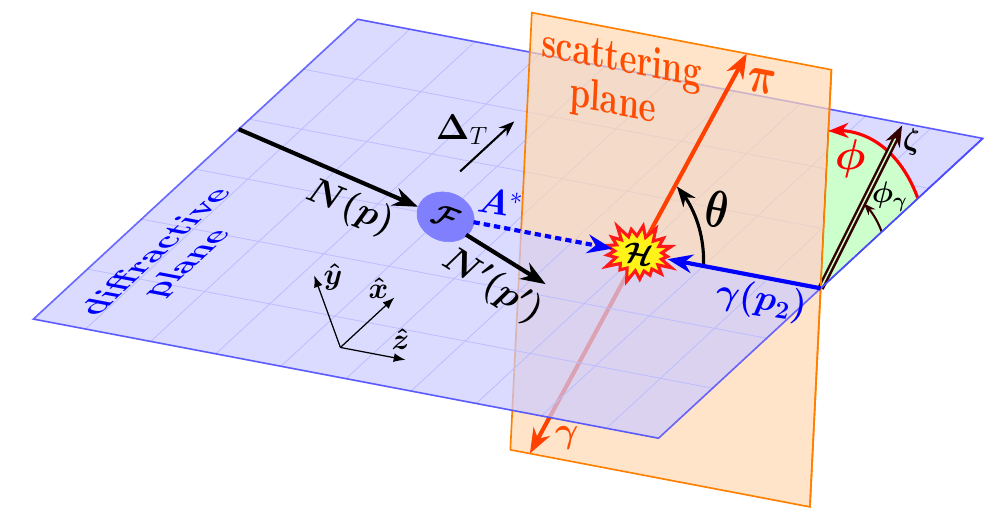}
		\includegraphics[scale=0.5]{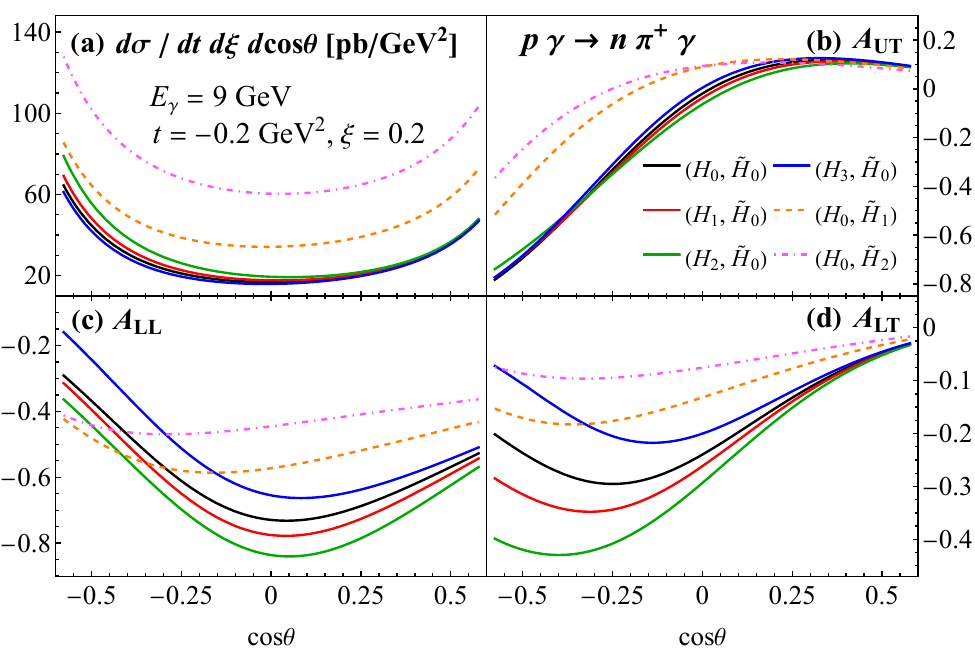}
	\caption{Left: SDHEP frame to analyze the process in \eq{eq:sdhep}.
		Right: Unpolarized cross section (a) and three polarization asymmetry (b)--(d) 
		distributions in terms of the polar angle $\theta$ of the produced pion, 
		evaluated from Eqs.~\eqref{eq:unpol-xsec}--\eqref{eq:asymmetries} using different GPD models.
	}
	\label{fig:photoproduction}
\end{figure}

The Hall-D photon beam and its energy and polarization are suitable for exclusive photoproduction of a photon-meson pair~\cite{Qiu:2023mrm, Qiu:2022pla, Duplancic:2023kwe, Duplancic:2022ffo, Duplancic:2018bum, Boussarie:2016qop},
\begin{equation}
	N(p) + \gamma(p_2) \to N'(p') + \pi(q_1) +  \gamma(q_2) \, ,
\label{eq:sdhep}
\end{equation}
which provides a new and advantageous observable to probe generalized parton distributions (GPDs) that
encode important tomographic parton images of nucleons~\cite{Burkardt:2000za, Burkardt:2002hr}.
In the kinematic region where the transverse momenta of $\pi(q_1)$ and $\gamma(q_2)$ are much greater than the invariant mass 
of the nucleon momentum transfer, $q_T \sim q_{1T} \sim q_{2T} \gg \sqrt{-t} \equiv \sqrt{-(p - p')^2}$, 
the scattering amplitude of the process in Eq.~\eqref{eq:sdhep} can be expressed in terms of nucleon GPDs with perturbatively calculable coefficients,
up to corrections in powers of $\sqrt{-t} / q_T$,
\begin{align}
	&\M_{ N \gamma_{\lambda} \to N' \pi \gamma_{\lambda'} }^{[F, \wt{F}]}(t, \xi, \theta, \phi)
	= \sum_{f,f'} e^{-i \lambda \phi} \cdot \int_{-1}^{1} dx \int_0^1 dz \, \bar{D}^{f'}_{\pi}(z)  \nn\\
		&\hspace{10em}\times
		\bb{ 
			F_{NN'}^f(x, \xi, t) \, \widetilde{C}^{ff'}_{\lambda\lambda'}(x, \xi; z; \theta)
			  + \wt{F}_{NN'}^f(x, \xi, t) \, C^{ff'}_{\lambda\lambda'}(x, \xi; z; \theta)
		},
\label{eq:factorize}
\end{align}
where $F$ and $\wt{F}$ are the GPDs associated with the nucleon diffraction from $N$ to $N'$,
$\bar{D}$ is the pion distribution amplitude, and $f$ and $f'$ represent the parton flavors.
The indices $\lambda$ and $\lambda'$ denote the photon helicities in the SDHEP frame~\cite{Qiu:2023mrm, Qiu:2022pla},
as shown in \fig{fig:photoproduction} (Left), which is the center-of-mass frame of the 
hard collision that produces the $\pi$ and $\gamma$ pair.

In \eq{eq:factorize}, the azimuthal angular $\phi$ dependence between the diffractive and scattering planes, 
as defined in Fig.~\ref{fig:photoproduction} (Left), was explicitly factored out of the hard coefficients $C$ and $\wt{C}$.
When the beam and/or target are polarized, this dependence helps to disentangle different GPDs.
In terms of the angles in \fig{fig:photoproduction} (Left), we can write the differential cross section of the process in Eq.~(\ref{eq:sdhep}) as~\cite{Qiu:2023mrm}
\begin{align}\label{eq:xsec}
	\frac{d\sigma}{d |t| \, d\xi \, d\cos\theta \, d\phi}
	&\, = \frac{1}{2\pi} \frac{d\sigma}{d |t| \, d\xi \, d\cos\theta }
		\cdot
		\big[ 1 +  \lambda_N \lambda_{\gamma} \, A_{LL}(t, \xi, \cos\theta) \nn\\
	&\;	+ \zeta \, A_{UT}(t, \xi, \cos\theta) \cos2(\phi - \phi_{\gamma})
			+ \lambda_N \, \zeta A_{LT}(t, \xi, \cos\theta) \sin2(\phi - \phi_{\gamma})
		\big],
\end{align}
where $\lambda_{\gamma}$ and $\lambda_N$ denote the longitudinal polarizations of the beam and target, respectively,
and $\zeta$ the linear polarization of the photon beam along the azimuthal direction $\phi_{\gamma}$.
\eq{eq:xsec} expresses the polarized differential cross section in terms of the unpolarized one,
\begin{align}\label{eq:unpol-xsec}
	\frac{d\sigma}{d |t| \, d\xi \, d\cos\theta }
	= \pi (\alpha_e \alpha_s)^2 \pp{ \frac{C_F}{N_c} }^2 \frac{1 - \xi^2}{\xi^2 s^3}\, \Sigma_{UU}(t, \xi, \cos\theta),
\end{align}
which is independent of $\phi$, with 
\beq[eq:sigma-UU]
	\Sigma_{UU}(t, \xi, \cos\theta)
	= | \M_{+}^{[\wt{H}]} |^2
		+ | \M_{-}^{[\wt{H}]} |^2
		+ | \wt{\M}_{+}^{[H]} |^2 
		+ | \wt{\M}_{-}^{[H]} |^2,
\eeq
and three polarization asymmetries that modulate the $\phi$ distribution,
\begin{align}
	A_{LL}(t, \xi, \cos\theta)
	&= 2 \, \Sigma_{UU}^{-1} \, \Re\bb{
					\M_{+}^{[\wt{H}]} \, \wt{\M}_{+}^{[H] *}
					+ \M_{-}^{[\wt{H}]} \, \wt{\M}_{-}^{[H] *}
		} , \nn\\
	A_{UT}(t, \xi, \cos\theta)
	&= 2 \, \Sigma_{UU}^{-1} \, \Re\bb{
					\wt{\M}_{+}^{[H]} \, \wt{\M}_{-}^{[H] *}						
					- \M_{+}^{[\wt{H}]}  \, \M_{-}^{[\wt{H}] *}
		} , \nn\\
	A_{LT}(t, \xi, \cos\theta)
	&= 2 \, \Sigma_{UU}^{-1} \, \Im\bb{
					\M_{+}^{[\wt{H}]} \, \wt{\M}_{-}^{[H] *}		
					+ \M_{-}^{[\wt{H}]} \, 	\wt{\M}_{+}^{[H] *}	
		} ,
\label{eq:asymmetries}
\end{align}
where the two subscripts of $\Sigma_{UU}$ and $A$'s refer to the target and beam polarizations, respectively, 
with $U$ for ``unpolarized'', $L$ for ``longitudinal polarized'', and $T$ for ``linearly polarized''.  
We have also reduced \eq{eq:factorize} into four independent helicity amplitudes $\M_{\pm}^{[\wt{H}]}$ and $\wt{\M}_{\pm}^{[H]}$,
which are convolutions of GPD $\wt{H}$ and $H$, respectively~\cite{Qiu:2023mrm}. 
Clearly, while the unpolarized cross section in \eq{eq:unpol-xsec} alone offers combined contributions of various GPDs,
the asymmetries in \eq{eq:asymmetries} provide independent observables to help disentangle contributions from different GPDs.
Especially, the $A_{LT}$ is directly given by the imaginary parts of the scattering amplitudes, 
which come from the ERBL region of GPDs for the process in \eq{eq:sdhep}, giving particular constraints on that region, 
including the $D$-term.

A special advantage of the process in \eq{eq:sdhep} is that the $\theta$ distribution is very sensitive to the $x$-dependence of GPDs,
contrary to many other processes including those photoproduction ones that can also be measured at Hall-D, 
such as the time-like Compton scattering (TCS)~\cite{Berger:2001xd, CLAS:2021lky} and 
diphoton photoproduction~\cite{Pedrak:2017cpp, Grocholski:2021man, Grocholski:2022rqj}.
As demonstrated in \fig{fig:photoproduction} (Right), starting from the base GPD models $(H_0, \wt{H}_0)$ 
(chosen as the GK model~\cite{Goloskokov:2005sd, Goloskokov:2007nt, Goloskokov:2009ia, Kroll:2012sm}),
adding different shadow GPD models~\cite{Bertone:2021yyz, Moffat:2023svr} one at a time results in 
different $\cos\theta$ distributions for both the unpolarized cross section and polarization asymmetries.
Notably, the GPD model $H_3$ differs from $H_0$ only in the ERBL region and affects $A_{LT}$ quite appreciably. 
On the other hand, contributions from these shadow GPDs would not be distinguished by the TCS 
and similar processes~\cite{Ji:1996nm, Radyushkin:1997ki, Brodsky:1994kf, Frankfurt:1995jw, Berger:2001zn, Goloskokov:2015zsa, Sawada:2016mao}.

\section{Polarized target for spectroscopy (Vincent Mathieu)}
Among the copious possible exclusive final states, the photoproduction of two pseudoscalar mesons on a nucleon target are one of the simplest and richest reactions~\cite{Mathieu:2019fts}. At least two mesons are required for meson spectroscopy and, for fixed (proton) target experiments, they will inevitably been produced with a recoiling nucleon. Since most pseudoscalar mesons are stable under the strong interaction, two of them in association with a nucleon form the simple final state for meson spectroscopy. However, at JLab energies $E_\text{lab}\sim~10$~GeV, the formation of nucleon resonance still represent an non-negligible background for meson spectroscopy. In both cases, the formation of a meson or a baryon resonance, the sub two-to-two process is described by several complex production amplitudes. Their complete determination required not only the knowledge of the differential cross section but also polarization observables. In this note, we will review which observables can be extracted with longitudinally and/or circularly polarized beam and a longitudinally polarized target. 

\begin{figure}[htbp]
	\centering
		\includegraphics[width=0.7\linewidth]{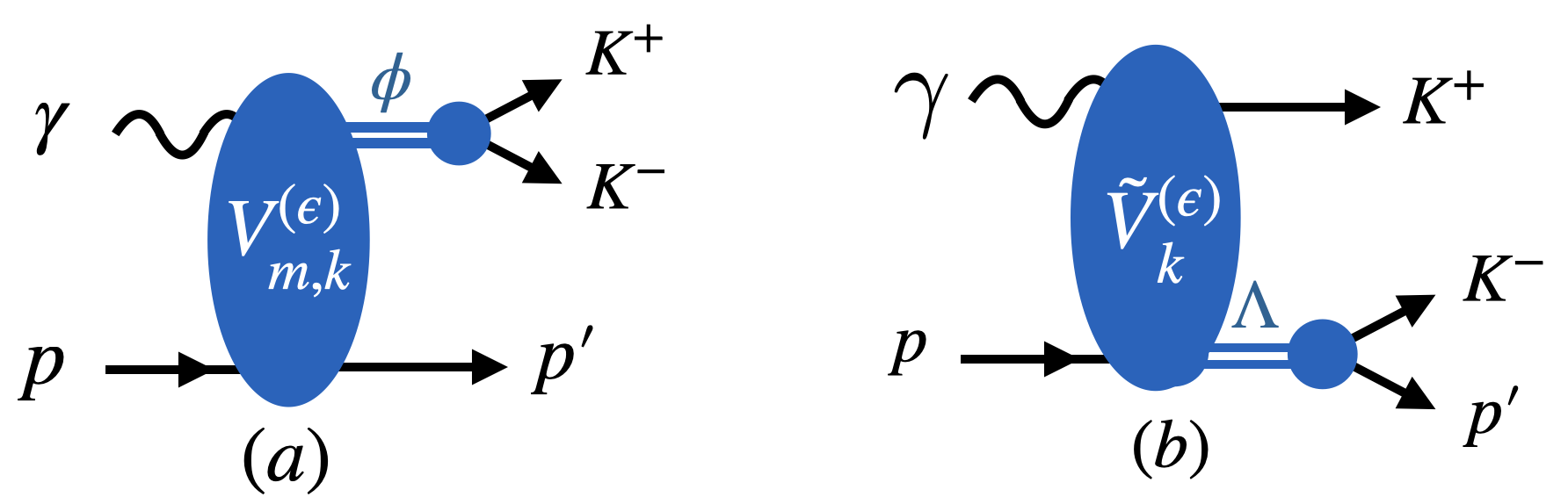}
	\caption{Examples of (a) meson resonance production and (b) baryon resonance production. }
	\label{fig:diagrams}
\end{figure}

In this note we will briefly review some observables accessible when the final state consist in two pseudoscalar mesons and one nucleon. An example of such reactions are $\gamma p \to K^+ K^- p$ illustrated on Fig.~\ref{fig:diagrams}. 
We thus consider a polarized beam and a polarized target. We do not assume any information about the polarization of the recoiling nucleon. The more general case involving a decay chain at the proton vertex is considered in Afzal's contribution in this document. In this case, the secondary baryonic decay leads to the extraction of the spin density matrix elements of the recoiling resonances and thus many more observables are accessible. 

The spin density matrices for both a real photon and a nucleon target can be decomposed using Pauli matrices
\begin{align}
    \rho^{\gamma/N} & =\frac{1}{2} I + \frac{1}{2} \vec P^{\gamma/N} \cdot \vec \sigma. 
\end{align}
We will consider a combination of linear and circular polarization for the beam. The target will be longitudinally polarized. Their polarization vector thus take the form
\begin{align}
    \vec P^\gamma(\Phi) & = (-\delta_\ell \cos 2 \Phi, -\delta_\ell \sin 2 \Phi, \delta_\odot),
    &
    \vec P^N & = (0,0,\delta_z)
\end{align}
$\delta_\ell, \delta_\odot$ are the degree of linear and circular polarization of the beam respectively. $\delta_z$ is the degree of target polarization. 

The intensity with a polarized photon beam and polarized target reads
\begin{align} \label{eq:intensity1}
    I(\Phi) & = \kappa \sum_{\lambda, \lambda'} \sum_{\lambda_1,\lambda'_1}
    \sum_{\lambda_2}
    A_{ \lambda, \lambda_1,\lambda_2} \ \rho^\gamma_{\lambda,\lambda'}(\Phi) \rho^N_{\lambda_1,\lambda'_1} \ 
    A^*_{ \lambda', \lambda_1',\lambda_2}
\end{align}
$\kappa$ is the phase space factor~\cite{Mathieu:2019fts}. The helicity amplitudes $A_{ \lambda, \lambda_1,\lambda_2}$ depend on five kinematic variables that will be specified later. In \eqref{eq:intensity1}, $\lambda,\lambda'$ are the photon helicities, $\lambda_1,\lambda'_1$ are the target helicities and $\lambda_2$ is the recoil helicity. Although the intensity $I(\Phi)$ is frame independent, the production amplitudes $A_{\lambda,\lambda_1,\lambda_2}$ depends on the frame via the definition of the helicities. 
This experimental set up leads to eight observables
\begin{align} \label{eq:intensity2}
    I(\Phi) & = \frac{\mathrm d \sigma}{\mathrm d t} 
    \left[(1+\delta_z P_z) - \delta_\ell(I^c + \delta_z P_z^c) \cos 2 \Phi  - \delta_\ell(I^s + \delta_z P_z^s) \sin 2 \Phi + \delta_\odot(I^\odot + \delta_z P_z^\odot) \right]
\end{align}
We have used the notation from~\cite{Roberts:2004mn}. All the observables depend on five kinematic variables. 

In many practical applications, we are interested in either the production of a meson resonance, or either a baryon resonance. These special cases are depicted in Fig.~\ref{fig:diagrams} for the process $\gamma(q_a, \lambda) p(q_b, \lambda_1) \to K^+ (q_1) K^-(q_2) p(q_3,\lambda_2)$. Let us investigate how the intensity \eqref{eq:intensity2} reduces in these two cases. 

We first perform a partial wave expansion in the two meson system. The five relevant variable for meson production are the total energy squared $s = (q_q+q_b)^2$, the momentum transferred between the two nucleons $u_3 = (q_b-q_3)^2$, the invariant mass squared of the two meson $s_{12} = (q_1+q_2)^2$ and the decay angles $\Omega_{12} = (\theta_{12}, \phi_{12})$ in the helicity frame.   We truncate the series to consider only one term, the one of spin $J$:
\begin{align}
    A_{\lambda,\lambda_1,\lambda_2} & = \sum_{m} V_{\lambda,m,\lambda_1,\lambda_2}(s,u_3) F(s_{12}) D^{J*}_{m,0}(\Omega_{12})
\end{align}
This applies, for instance when the invariance $K^+K^-$ mass is around 1 GeV. The partial wave expansion is largely dominating by the $J=1$ meson $\phi$, whose lineshape is encoded in the function $F(s_{12})$. 

The photon helicity $\lambda$ is replaced by the reflectivity $\epsilon$ in a standard way~\cite{Mathieu:2019fts} and parity invariance is used to reduce the number of production amplitude $V^{(\epsilon)}_{m,k}(s,u_3)$ with $k=0,1$ labelling the nucleon transitions (helicity non-flip and flip respectively). Parity invariance and the restriction of a single partial waves of spin $J$ implies $P_z = P^c_z = P^s_z = I^\odot = 0$ when the decay angles $\Omega_{12}$ are integrated over. The non-zero observables take the form
\begin{subequations}\begin{align}
    \frac{\mathrm d \sigma}{ \mathrm d t} 
    & = \frac{\kappa}{2} \sum_{m,k}  | V^{(+)}_{m,k}|^2 + |V^{(-)}_{m,k}|^2
    &
     I^c \frac{\mathrm d \sigma}{\mathrm d t}  & = \kappa \sum_{m,k} \operatorname{Re} \left[ V^{(+)}_{m,k} V^{(+)*}_{-m,k}  - V^{(-)}_{m,k} V^{(-)*}_{-m,k}  \right]
    \\
    P^\odot_z \frac{\mathrm d \sigma}{ \mathrm d t} 
    & = \kappa \sum_{m,k} \operatorname{Re} \left[V^{(+)}_{m,k} V^{(-)*}_{m,k}  \right]
    &
    I^s \frac{\mathrm d \sigma}{\mathrm d t}  & = \kappa \sum_{m,k} (-1)^m \operatorname{Im} \left[ V^{(+)}_{m,k} V^{(+)*}_{-m,k}  - V^{(-)}_{m,k} V^{(-)*}_{-m,k}  \right]
\end{align}
\label{eq:meson}\end{subequations}
The integration over the decay angles leads to four real observables to determine $2\times2\times (2J+1)$ complex production amplitudes $ V^{(\epsilon)}_{m,k}$.\footnote{The phase of one of each reflectivity component is arbitrary.} when the angular distribution of the meson pair is analyzed, several quadratic combination of $V^{(\epsilon)}_{m,k} V^{(\epsilon)}*_{m',k}$ are accessible. The general description of analysis with a linearly polarized beam is provided in Ref.~\cite{Mathieu:2019fts}. That is, the moments expansion of cross section and the observables $I^{c,s}(\Omega_{12})$. With a polarized target, one will have access to the moment expansion of the observable $P^\odot_z(\Omega_{12})$. As suggested by their integrated version in \eqref{eq:meson}, the polarized target observables is the only one probing the relative phase between both naturality components. The other observables are incoherent sums over the reflectivity. 

\medskip
Let us now consider the production of a baryon resonance as depicted in panel $(b)$ of Fig.~\ref{fig:diagrams}. We perform a partial wave expansion in the $(23)$ channel (particle 3 being the recoiling nucleon). The relevant variables are $s$, the momentum transferred at the photon vertex $t_1 = (q_a-q_1)^2$, the invariant mass squared $s_{23} = (q_2+q_3)^2$ and the two decay angles $\Omega_{23}$ of the recoiling nucleon in the rest frame of the resonance:
\begin{align}
    A_{\lambda,\lambda_1,\lambda_2} & = \sum_{m} \tilde V_{\lambda,m,\lambda_1,\lambda_2}(s,t_1) \tilde F_{\lambda_2}(s_{23}) D^{\tilde J*}_{m,\lambda_2}(\Omega_{23})
\end{align}
It worth noting that, thanks to the spin of the final nucleon, the partial wave of spin $\tilde J$ includes both parity states. Selecting a good parity states imposes further constraints on the lineshape $\tilde F_{\lambda_2}(s_{23})$. 
The reflectivity is introduced such that it matches the naturality of the exchange in the high energy limit
\begin{align}
    \tilde V^{(\epsilon)}_{\lambda_1,m}(s,t) &=   \frac{1}{2} \left[ \tilde V_{+1,\lambda_1,m}(s,t)  + \epsilon \tilde V_{-1,\lambda_1,m}(s,t) \right]
\end{align}
Since they satisfy the parity relations $\tilde V^{(\epsilon)}_{-\lambda_1,-m}(s,t)  = \tau \epsilon (-1)^{m-\lambda_1}\tilde V^{(\epsilon)}_{\lambda_1,m}(s,t) $, with $\tilde \tau = \tilde P(-1)^{\tilde J-1/2}$ the naturality of the baryon resonance having spin-parity $\tilde J^{\tilde P}$ in the channel $(23)$, there are only $2\tilde J+1$ independent transition amplitudes $\tilde V^{(\epsilon)}_k\equiv \tilde V^{(\epsilon)}_{\frac{1}{2}, k}$ with $k = -J,\ldots, +J$. 

We are ready to particularize the observables for the case of a single baryon resonance produced. 
We note that only three observables survive. In addition to the usual cross-section and beam asymmetry (obtain from the polarization of the beam), the polarized target provide a third possible measurement:
\begin{subequations}\begin{align}
    \frac{\mathrm d \sigma}{ \mathrm d t} 
    & = \kappa \sum_{k}  |\tilde V^{(+)}_{k}|^2 + |\tilde V^{(-)}_{k}|^2
    &
     I^c \frac{\mathrm d \sigma}{\mathrm d t}  & = \kappa \sum_k \operatorname{Re} |\tilde V^{(+)}_{k} |^2   -|\tilde V^{(-)}_{k} |^2
    \\
    P^\odot_z \frac{\mathrm d \sigma}{ \mathrm d t} 
    & = \kappa \sum_{k} \operatorname{Re} \left[\tilde V^{(+)}_{k} \tilde V^{(-)*}_{k}  \right]
    &
    P^s_z & = \kappa \sum_{k} \operatorname{Im} \left[\tilde V^{(+)}_{k} \tilde V^{(-)*}_{k}  \right]
\end{align}\end{subequations}
As in the case of a meson resonance production, a polarized target will provide information about the interference between the two reflectivity components.

\section{Hyperon spectroscopy prospects with a polarized target (Farah Afzal)}
As mentioned in Sec.~\ref{sec:REGGE}, the PAC48 approved the measurement of the high-energy contribution to the GDH sum rule~\cite{Dalton:2020wdv} in Hall D, which requires a longitudinally polarized target. The possibility to conduct measurements with polarized targets in Hall D provides new opportunities to explore the hadron spectra, and in particular study excited baryons. This will shed light on the strong force and provide valuable information on the relevant degrees of freedom for quantum chromodynamics (QCD).

Taking into account quark model~\cite{Ronniger:2011td} and lattice QCD~\cite{Edwards:2012fx} calculations, a lot more excited hyperons, e.g. $\Lambda^*$ and $\Sigma^*$ states, should exist than there are currently listed in the PDG~\cite{ParticleDataGroup:2022pth}. The existing data base in the field of hyperon spectroscopy is rather sparse and is based mostly on $K^-$ induced reactions from the 1970s~\cite{ParticleDataGroup:2022pth}. First partial-wave analyses (PWAs) of the existing data shows that there are discrepancies between different PWAs regarding the properties of many states. In addition, many one- and two-star rated resonances, as well as some newly suggested resonances need confirmation. An overview is given in Ref.~\cite{Klempt:2020bdu}. The planned experiments in Hall D can play a major role in providing much needed new data by probing excited hyperons with two different production mechanisms, with a $K_L$ beam with the already approved KLong Facility (KLF)~\cite{KLF:2020gai}, and by using a photon beam. The data will provide the opportunity to potentially discover \textit{missing resonances}, and study the nature and properties of not-well established resonances.\\

\begin{figure}[ht]
    \centering
    \includegraphics[width=0.5\textwidth]{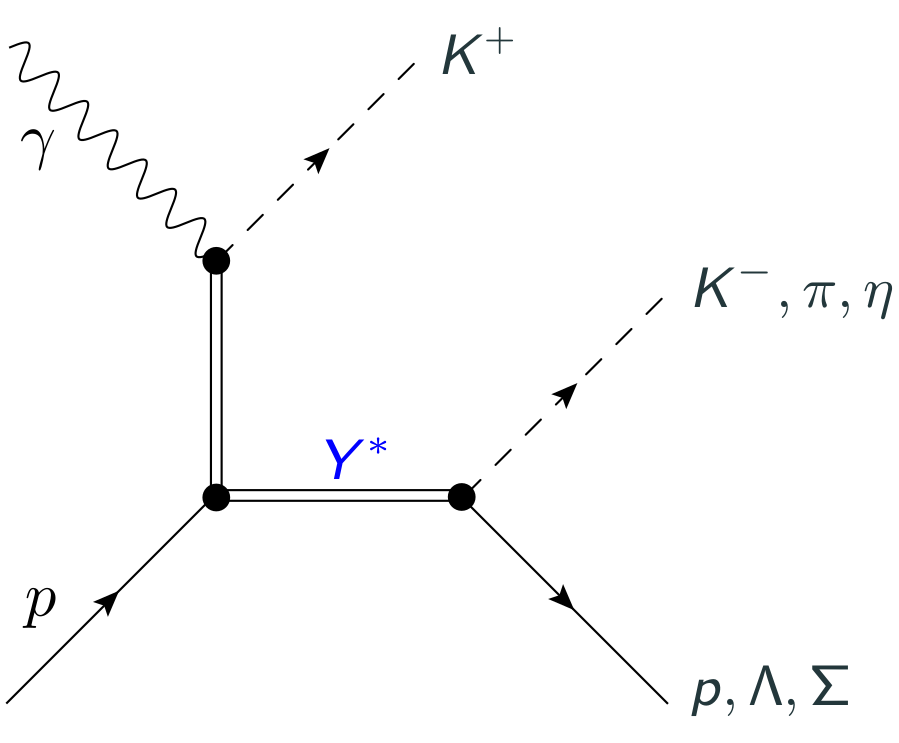}
    \caption{Production of excited hyperons ($Y^*$) via t-channel exchange. Possible exchanged particles are for example $K,K^*,K_2^*$. Taken from~\cite{LOIhyperons}.}
    \label{fig:t-channel}
\end{figure}

In the Letter-of-Intent (LOI 12-24-001)~\cite{LOIhyperons} submitted to the PAC52, we propose to investigate the excited $\Lambda^*$ and $\Sigma^*$ spectra by studying several different photo-induced reactions using an elliptically polarized photon beam~\cite{A2:2024ydg} and a longitudinally polarized target with the GlueX experiment in Hall D.
Fig.~\ref{fig:t-channel} shows the dominant $t$-channel production of excited hyperons $Y^*$ in photo-induced reactions, where the excited hyperons can decay e.g. via $pK^-$, $\Lambda\pi^0$ and $\Lambda\eta$. Considering the spins of the initial and final state particles, these 3-body final states can be described by eight complex amplitudes. For an unambiguous solution at least 16 out of possible 64 polarization observables need to be measured~\cite{Roberts:2004mn,Kroenert:2020ahf}. These polarization observables are accessible using a polarized photon beam, a polarized target, and measuring the polarization of the recoiling baryon. 

As outlined in Sec.~\ref{sec:circpol}, an elliptically polarized photon beam with both a linear and circular polarization component will be available at the GlueX experiment in the future. Combining it with a longitudinally polarized target and making use of the self-analyzing power of the ground state hyperons $\Lambda$ and $\Sigma$, it will be possible to measure the following set of 32 polarization observables given in the notation used in Ref.~\cite{Roberts:2004mn}:  
\begin{align}
 &\Big\{ I_{0}, \hspace*{0.5pt}  I^{s}, \hspace*{0.5pt} I^{c}, \hspace*{0.5pt} I^{\odot}, \hspace*{0.5pt} P^{s}_{z}, \hspace*{0.5pt} P^{c}_{z}, \hspace*{0.5pt} P^{\odot}_{z}, \hspace*{0.5pt} P_{z}, \hspace*{0.5pt}P_{x'}, \hspace*{0.5pt} P_{y'}, \hspace*{0.5pt} P_{z'}, \hspace*{0.5pt} P^{\odot}_{x'}, \hspace*{0.5pt} P^{\odot}_{y'}, \hspace*{0.5pt} P^{\odot}_{z'}, \hspace*{0.5pt} P^{s}_{x'}, \hspace*{0.5pt} P^{s}_{y'}, \hspace*{0.5pt} P^{s}_{z'}, \hspace*{0.5pt}  P^{c}_{x'}, \hspace*{0.5pt} P^{c}_{y'}, \hspace*{0.5pt} P^{c}_{z'}, \hspace*{0.5pt} \nonumber \\
   & \hspace*{6.75pt}  \Ocal_{zx'}, \hspace*{0.5pt} \Ocal_{zy'}, \hspace*{0.5pt} \Ocal_{zz'}, \hspace*{0.5pt} \Ocal^{\odot}_{zx'}, \hspace*{0.5pt} \Ocal^{\odot}_{zy'}, \hspace*{0.5pt} \Ocal^{\odot}_{zz'}, \hspace*{0.5pt} \Ocal^{s}_{zx'}, \hspace*{0.5pt} \Ocal^{s}_{zy'}, \hspace*{0.5pt} \Ocal^{s}_{zz'}, \hspace*{0.5pt} \Ocal^{c}_{zx'}, \hspace*{0.5pt} \Ocal^{c}_{zy'}, \hspace*{0.5pt} \Ocal^{c}_{zz'} \Big\} . \label{eq:All32ObservablesPWA}
\end{align}  

In a truncated PWA with truncation-orders~$J_{\text{max}}$ up to $5/2$ or $7/2$, we expect this set of observables to result in a unique solution. In particular, the following polarization observables requiring a polarized target $P_z, O_{zz'}, P_z^\odot, \Ocal_{zz'}^\odot$ are essential for extracting a unique solution for the partial waves~\cite{LOIhyperons}. \\
The proposed experiment would use the 12~GeV CEBAF beam energy with a beam current of around 100~nA, a coherent edge position of around $6.8$~GeV, a longitudinally polarized butanol target with the parameters summarized in Tab.~\ref{tab:lumi} and the standard GlueX detector system. An estimated data-taking period of about 100 PAC days should enable us to significantly contribute towards achieving a complete experiment and explore the hyperon spectra. Clear signals for well known hyperons e.g. $\Lambda(1520)$ are visible in the GlueX-I data set. More details are given in~\cite{LOIhyperons}.

\begin{figure}[h!]
    \begin{overpic}[width=0.6\textwidth]{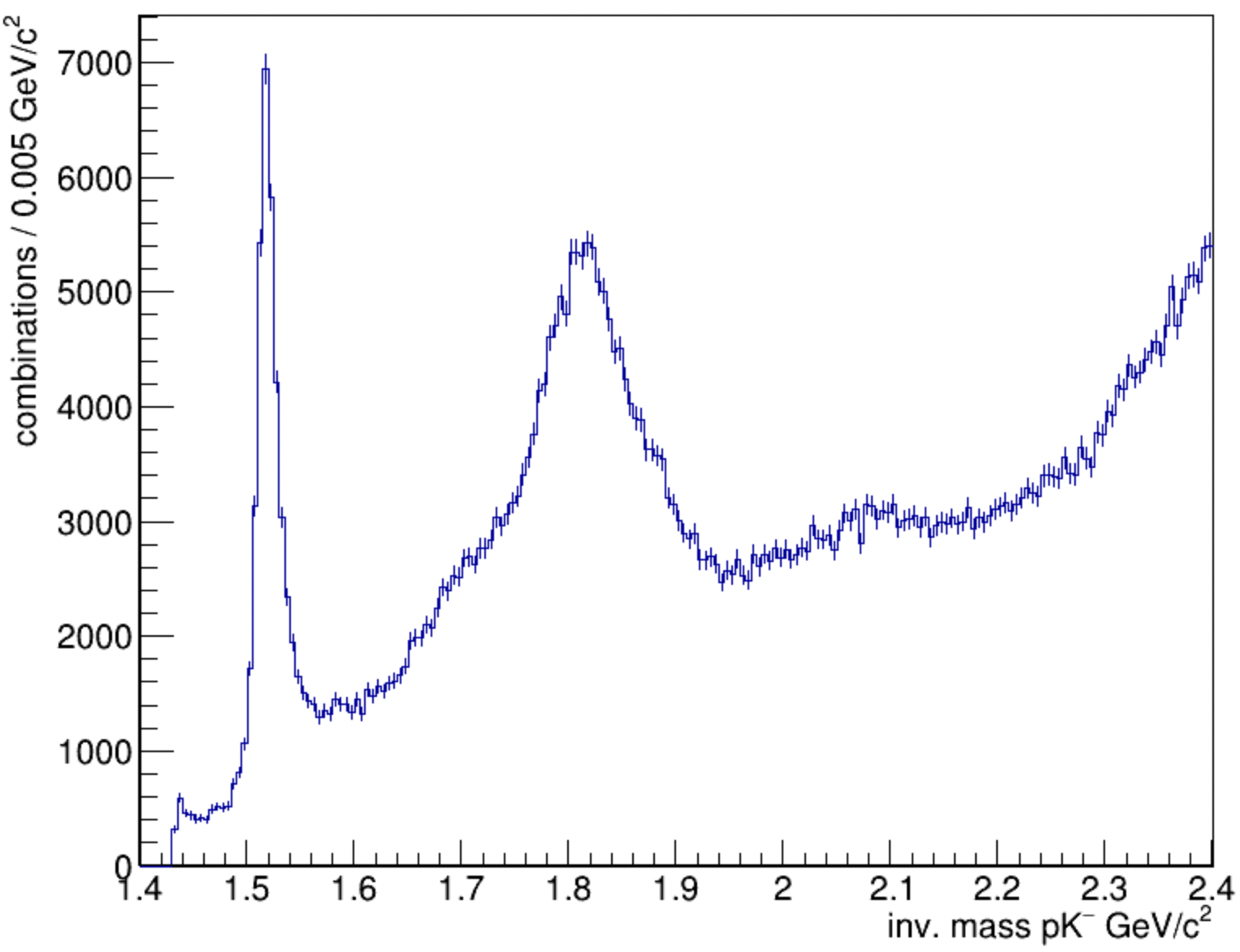}
\put(24,70){\tiny{$\boldsymbol{\Lambda(1520)\frac{3}{2}^-}$}}
\put(22,15){\tiny{$\boldsymbol{\Lambda(1600)\frac{1}{2}^+}$}}
\put(26,55){\tiny{$\boldsymbol{\Sigma(1660)\frac{1}{2}^+}$}}
\put(26,51){\tiny{$\boldsymbol{\Sigma(1670)\frac{3}{2}^-}$}}
\put(26,47){\tiny{$\boldsymbol{\Lambda(1670)\frac{1}{2}^-}$}}
\put(26,43){\tiny{$\boldsymbol{\Lambda(1690)\frac{3}{2}^-}$}}
\put(26,39){\tiny{$\textcolor{gray}{\Lambda(1710)\frac{1}{2}^+}$}}

\put(38,30){\tiny{$\boldsymbol{\Sigma(1750)\frac{1}{2}^-}$}}
\put(38,26){\tiny{$\boldsymbol{\Sigma(1775)\frac{5}{2}^-}$}}

\put(41,20){\tiny{$\boldsymbol{\Lambda(1800)\frac{1}{2}^-}$}}
\put(41,16){\tiny{$\boldsymbol{\Lambda(1810)\frac{1}{2}^+}$}}
\put(41,12){\tiny{$\boldsymbol{\Lambda(1820)\frac{5}{2}^+}$}}
\put(41,8){\tiny{$\boldsymbol{\Lambda(1830)\frac{5}{2}^-}$}}

\put(55,50){\tiny{$\boldsymbol{\Sigma(1910)\frac{3}{2}^-}$}}
\put(55,46){\tiny{$\boldsymbol{\Sigma(1915)\frac{5}{2}^+}$}}
\put(55,42){\tiny{$\boldsymbol{\Lambda(1890)\frac{3}{2}^+}$}}

\put(60,27){\tiny{$\textcolor{gray}{\Lambda(2000)\frac{1}{2}^-}$}}
\put(60,23){\tiny{$\textcolor{gray}{\Lambda(2050)\frac{3}{2}^-}$}}
\put(60,19){\tiny{$\textcolor{gray}{\Lambda(2070)\frac{3}{2}^+}$}}
\put(60,15){\tiny{$\textcolor{gray}{\Lambda(2080)\frac{5}{2}^-}$}}
\put(60,11){\tiny{$\textcolor{gray}{\Lambda(2085)\frac{7}{2}^+}$}}

\put(78,24){\tiny{$\boldsymbol{\Sigma(2030)\frac{7}{2}^+}$}}
\put(78,20){\tiny{$\boldsymbol{\Lambda(2100)\frac{7}{2}^-}$}}
\put(78,16){\tiny{$\boldsymbol{\Lambda(2110)\frac{5}{2}^+}$}}
\end{overpic}
    \caption{The invariant mass of $K^-p$ is shown for the incoming photon energy range of the coherent peak region between 8.2-8.8~GeV, which corresponds to about $20$\% of the total GlueX-I data. $3*$ and $4*$ resonances are shown in bold, while $1*$ and $2*$ resonances are shown in gray~\cite{ParticleDataGroup:2022pth}. Taken from~\cite{LOIhyperons}.}
    \label{fig:Kmp}
\end{figure}
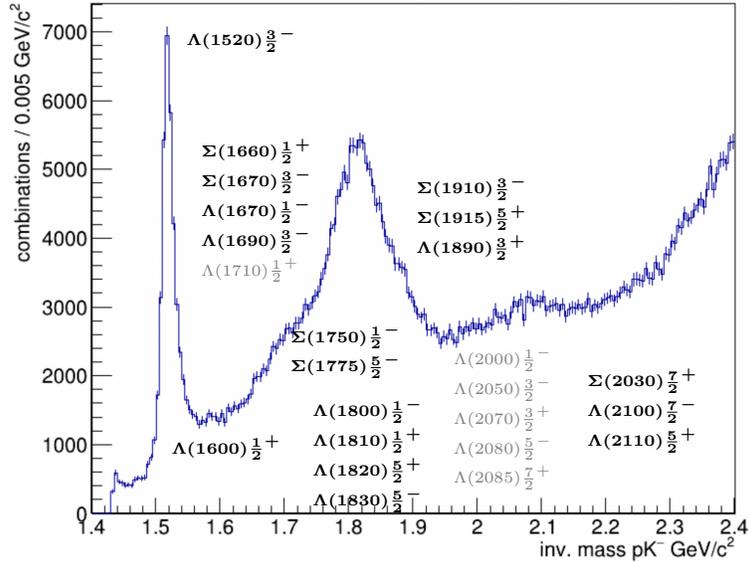

\section{Summary}

Hall D can be equipped with a polarized target that can achieve a very high polarization for multiple potential target species.
This target is already approved to run the REGGE experiment, which will require polarized hydrogen and  deuterium.  This experiment will measure the high energy contribution to the GDH sum rule for the proton and the neutron.
The ability to significantly polarize other species, such as $^7$Li, $^{13}$C, $^{17}$O and $^{19}$F opens up the possibility to measure the medium modifications of $\kappa/M$ for a nucleon in a nucleus through the measurement of the GDH integrand for the embedded nucleon.
The target is also capable of high negative tensor polarization in deuterium which will present opportunities not covered in this white paper.
A circularly polarized photon beam is also required for many of the observables mentioned in this document. This has already been demonstrated in Hall D, and the first physics analyses with that data are underway.
Measuring GPDs with a photon beam and polarized target would allow measurement of the $x$-dependence of GPDs, help disentangle contributions from different GPDs and provide sensitivity to shadow GPDs in a way not available with other measurements.
The availability of a polarized target will also allow us to access information about the interference between the two reflectivity components for both meson and baryon spectroscopy.  
Baryon spectroscopy of hyperons which have a self analyzing decay would allow the measurement of observables $P_z, O_{zz'}, P_z^\odot, \Ocal_{zz'}^\odot$, which are needed for a ``complete experiment" allowing for the extraction of the partial wave amplitudes without ambiguities.

\acknowledgments
We gratefully acknowledge the contributions of Sean Dobbs, Justin Stevens, and Matthew Shepherd in the conception, organization and running of the workshop.
This material is based upon work supported by the U.S. Department of Energy (DOE), Office of Science, Office of Nuclear Physics under contract DE-AC05-06OR23177 and the UK Science and Technology Facilities Council.
V.M. is a Serra Húnter fellow and acknowledges support from the Spanish national Grants PID2020-118758GB-I00 and CNS2022-136085.

\bibliography{references}

\end{document}